# First-principles study of electronic structures and optical properties of Cu, Ag, and Au-doped anatase TiO$_2$

Meili Guo and Jiulin Du[*]

*Department of Physics, School of Science, Tianjin University, Tianjin 300072, China*

**Abstract** We perform first-principles calculations to investigate the band structure, density of states, optical absorption, and the imaginary part of dielectric function of Cu, Ag, and Au-doped anatase TiO$_2$ in 72 atoms systems. The electronic structure results show that the Cu incorporation can lead to the enhancement of *d* states near the uppermost of valence band, while the Ag and Au doping cause some new electronic states in band gap of TiO$_2$. Meanwhile, it is found that the visible optical absorptions of Cu, Ag, and Au-doped TiO$_2$, are observed by analyzing the results of optical properties,.which locate in the region of 400-1000 nm. The absorption band edges of Cu, Ag, and Au-doped TiO$_2$ shift to the long wavelength region compared with the pure TiO$_2$. Furthermore, according to the calculated results, we propose the optical transition mechanisms of Cu, Ag, and Au-doped TiO$_2$, respectively. Our results show that the visible light response of TiO$_2$ can be modulated by substitutional doping of Cu, Ag, and Au.

## 1. Introduction

TiO$_2$ has several good properties, such as nontoxic, relatively inexpensive, physical and chemical stability, and high photocatalytic activity [1]. Because of these excellent properties, TiO$_2$ has attracted extensive attentions during last decades as an extremely promising photocatalyst for environmental applications. However, TiO$_2$ has a wide band gap, which limits the application of visible light [2]. Therefore, the effective utilization of visible light is one of the important subjects for TiO$_2$ as a photocatalyst. Enormous efforts have been devoted to extend the light absorption range of TiO$_2$ from UV to visible light, such as impurity doping or surface modifications. Particularly, impurity doping has been widely researched by chemical synthesis and ion implantation methods in recent years [3, 4].

---
[*]*E-mail address:* jiulindu@yahoo.com.cn.

The visible photoactivities induced by Ag and Au doping into the lattice of TiO$_2$ have been reported in recent years. The first principles calculations of electronic structures of the Ag-doped anatase TiO$_2$ were carried out, and the researchers supposed that the visible light absorption was due to the Ag 4$d$ states mixed with Ti 3$d$ states in the band gap. In addition, Ag was doped and deposited on TiO$_2$ simultaneously by experiment, which was more effective than the single deposition, because Ag and Ag$^+$ could both act as electron traps to enhance photocatalytic activity [5-7]. The results of Liu *et al.* showed the enhanced photocatalytic activity of Ag-doped TiO$_2$ particles, because the structure modification of TiO$_2$ resulted by Ag doping could facilitate the produce of O vacancy [8]. Dombi *et al.* also reported that Ag$^+$ could act as a efficient electron scavenger, and yield a significantly enhanced photodegradation rate [9]. Similar results were found in Au/Au$^{3+}$-TiO$_2$ photocatalysts toward visible photooxidation for wastewater treatment. They proposed that the Au$^+$ doped into the lattice of TiO$_2$ acted as electron trap, which could promote the charge trapping and favor the charge migration to O$_2$, and thus increase the photoactivity [10].

The Cu-doped TiO$_2$ was discussed relatively more than the Ag and Au-doped TiO$_2$ [11-18]. Wu *et al.* reported that the Cu particles deposited on TiO$_2$ could lead to significant enhancement of photocatalytic activity for H$_2$ production from aqueous methanol solution. On the contrary, the doping of Cu ion into TiO$_2$ lattice would result in reduction of photocatalytic activity [11]. Park *et al.* carried out the research about the photocatalytic activity of 2.5 wt% Cu-doped TiO$_2$ nanopowders which exhibited an absorption edge at 480–490 nm longer than that of rutile TiO$_2$ powder, and led to a two times faster degradation of 4-cholorophenol [12]. Similar conclusion was reported that the Cu doping was effective for the visible-light photocatalysis of TiO$_2$ films, and the charge separation of the photogenerated electron/hole pairs took place effectively in the Cu-doped TiO$_2$ film [13]. However, these results are contradictory. To the best of our knowledge, no systematic research and explanation on optical transition mechanism of Cu, Ag, and Au-doped TiO$_2$ have been presented. Up to now, lots of optical transition mechanism can only be deduced by the electronic structure, which is difficult to reveal the real photocatalytic mechanism.

In previous works, we explored the electronic structure and visible photoactivity of B, Cd, and F-doped $TiO_2$ [19-21]. In present study, we calculate the electronic structure and optical transition of Cu, Ag, and Au-doped $TiO_2$ in detail. The paper is organized as follows: in section 2, we describe the basic ingredients and details of computational methods. Then, in section 3, the results of our calculations are presented and discussed. First, we calculate the electronic structures, because the optical properties depend on both the interband and intraband transitions, which are determined by energy band. Second, we analyze the optical absorption of different doping systems. Third, the optical transition mechanisms of Cu, Ag, and Au-doped $TiO_2$ have been discussed. Finally, the conclusion is given in section 4.

**2. Calculation Methods**

Cambridge Serial Total Energy Package (CASTEP) has been used in our calculations, which is based on density functional theory (DFT) using a plane-wave pseudopotential method [22]. We use the generalized gradient approximation (GGA) in the scheme of Perdew-Burke-Ernzerhof (PBE) to describe the exchange-correlation functional [23]. The interaction between valence electrons and the ionic core is described by ultrasoft pseudopotential [24], which is used with $2s^22p^4$, $3d^24s^2$, $3d^{10}4s^1$, $4d^{10}5s^1$, and $5d^{10}6s^1$ as the valence electrons configuration for the O, Ti, Cu, Ag, and Au atoms, respectively. The nonlinear core correction (NLCC), proposed by Louie *et al.*[25], has been used for Ti, Cu, Ag, and Au ions. It has been shown that the NLCC term is especially important when semicore states are not explicitly treated as valences.

In addition, because the previous works showed that the relativistic effects were non-negligible for the $3d$ states of transition metals, important for $4d$ states of transition metals, and so important for $5d$ states of transition metals [26, 27], we evaluate the relativistic effects on electronic structures of doping elements. All electronic relativistic option is carried out by using $Dmol^3$ codes. Except for the valence elctrons, the core electrons are also included in the calculation [28, 29].

The calculations are performed at consistent volume, and the $TiO_2$ supercell has been used, which contains 72 atoms. The substitutional method has been taken into account in this paper. The Cu, Ag, and Au atoms are used to substitute Ti atom in $TiO_2$, because atom radius of these metals are larger than the O atom. Recent investigation indicated that the distance of doping atoms could influence the calculated results due to the interaction of doping atoms [30]. Therefore, one dopant atom is only contained in one supercell to weak the interaction with dopant atoms. In this case, the doped $TiO_2$ systems form the configurations of $Ti_{23}CuO_{48}$, $Ti_{23}AgO_{48}$, and $Ti_{23}AuO_{48}$, corresponding to the concentration of 4.17%. The calculated structure of substitutional doping system is presented in Fig.1. We choose the energy cutoff to be 340 eV for the pure, Cu, Ag, and Au-doped $TiO_2$. The Brillouin-zone sampling mesh parameters for the $k$-point set are 3×2×3 for 72 atoms systems. Energy convergence test indicates the stability of systems. And then, the positions of substitutional atoms and lattice constants of the doped $TiO_2$ systems are optimized. In the optimization process, the energy change, maximum force, maximum stress and maximum displacement tolerances are set as $1\times10^{-5}$ eV/atom, 0.03 eV/Å, 0.05 Ga, and 0.001 Å, respectively. The lattice constants of pure $TiO_2$ are 3.776 Å for $a$ and 9.4860 Å for $c$, respectively, which are consistent with the experimental data [31]. Besides, for obtaining exact optical absorption spectra in the low energy range, the scissors operation has been carried out in optical absorption of $TiO_2$.

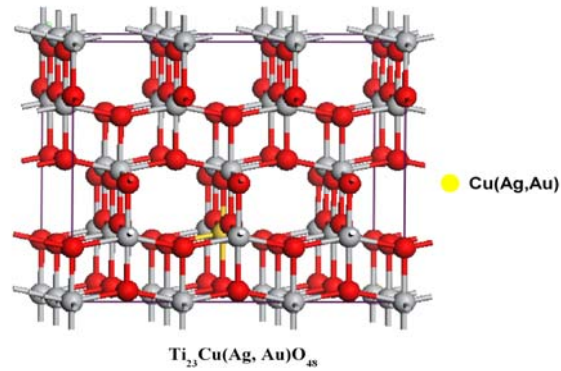

Fig.1 Configuration of $Ti_{23}$Cu (Ag, Au) $O_{48}$, corresponding to 72 atoms.

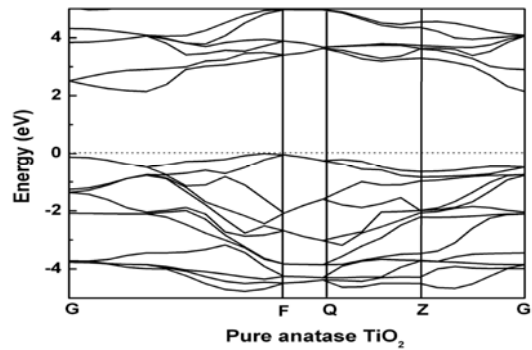

Fig.2 Band structure of pure $TiO_2$.

## 3. Results and Discussions

### 3.1 Band structure and DOS of pure and doped anatase $TiO_2$.

Fig.2 shows the band structure of pure anatase $TiO_2$. It is observed that the band gap is about 2.18 eV, which is less than the experimental result of 3.2 eV. It is well known that the underestimated band gap can be due to the choice of exchange-correlation energy. The top of valence band approximately locates near the F-point and the bottom of the conduction band locates at the G-point, which means that $TiO_2$ with anatase structure is an indirect-gap material. The valence band of pure $TiO_2$ mainly consists of the $2p$, $2s$ states of O and $3d$ states of Ti. In the uppermost valence band, the O $2p$ states are predominantly found between -5 and 0 eV, while the O $2s$ states appear in the range from -18 to -15.5 eV. The Ti $3d$ states give rise to some bands in the energy range from -5 to -3 eV. The lowest conduction band is dominated by Ti $3d$ states. Meanwhile, the hybridization between the Ti $3d$ and O $2p$ levels at the valence band can be observed, which is very close to the previous calculated results [32]. In addition, we also calculate the effect of NLCC on electronic structure, and find that the deviation of band gap is less than 0.1 eV. For computational methods, it should be pointed out that LDA (GGA) +U can correct the energy level of Ti $3d$ states by adding the Hubbard-U to Ti $d$ states, and the band gap can be corrected to 2.8 eV [33, 34]. The hybrid functional, such as PBE0 and HSE, can correct the Ti $3d$ states, and the band gap of $TiO_2$ is 3.2 eV, very close to the experimental data of 3.3 eV, but it is computationally expensive [32].

To obtain the outline of electronic structures of Cu, Ag, and Au-doped $TiO_2$, we perform the density of states (DOS) of doped $TiO_2$ and partial density of states (PDOS) of dopants in the configuration of $Ti_{23}Cu (Ag, Au) O_{48}$, which are presented in Fig.3 (a) and Fig.3 (b), respectively. Here, the NLCC is only considered as a way of obtaining accurate pseudopotential descriptions of heavy-metal doping systems. Indeed, it has been shown that Cu, Ag and Au adsorbed $TiO_2$ can obtain relatively high accuracy and the NLCC is very important for spin non-polarized systems with semicore electrons [35]. Thus, we consider the NLCC in all calculations. Besides, we consider the relativistic effects on Cu, Ag, and Au-doped $TiO_2$ systems, because the relativistic effects play an important role on heavy elements. However, we find that the relativistic effects are not obvious for Cu, Ag and Au electronic states, which may be related to the low concentration of doping metal on $TiO_2$. Therefore, the electronic structures and optical calculations of Cu, Ag and Au-doped TiO$_2$, based on relativistic effects, have not been further considered. Compared with the pure TiO$_2$, the strong *d* electronic states are observed in different energy ranges after doping metals in Fig.3 (a). The Ag and Au doping can induce the middle states in band gap, which are mainly consisted of the *d* and *p* electronic states. From the results of PDOS of Ag and Au atoms in Fig.3 (b), the *d* electronic states are mainly attributed to the Ag and Au, while the *p* electronic states are due to the O 2*p* states. These results are consistent with the previous experimental and theoretical results [5, 6, 10, 36]. In early calculations, the Ag-doped TiO$_2$ with high concentration (6.25 %) in 48 atoms system also produced middle states, and it was 0.65 eV above the valence band and 0.76 eV below the conduction band [5]. These midgap electronic states may induce some visible optical transition. Meanwhile, the *p* and *d* states have been observed in the energy range of 7-9 eV at the bottom of valence band due to the doping of Ag and Au. These *p* and *d* states are similar to the midgap states. To the Cu-doped TiO$_2$, it is observed that the 3*d* states of Cu mainly appear at the valence band, particularly, dominated at the top and bottom of valence band in Fig.3 (b). Thus, the enhanced *d* states of Cu-doped $TiO_2$ at the uppermost valence band are due to the Cu dopant, which is consistent with the previous theoretical results [14, 15]. Even in the high concentration doping ($Ti_3CuO_7$) system, the Cu 3*d* states also located above the valence band using full-potential linearized-augmented-plane wave method [15]. The increasing *d* electronic states of Cu-doped $TiO_2$ at the uppermost valence band lead to the band gap narrowing. Meanwhile, the Ti 3*d* states decrease slightly for Cu, Ag, and Au-doped $TiO_2$ due to the substitution of metal ions.

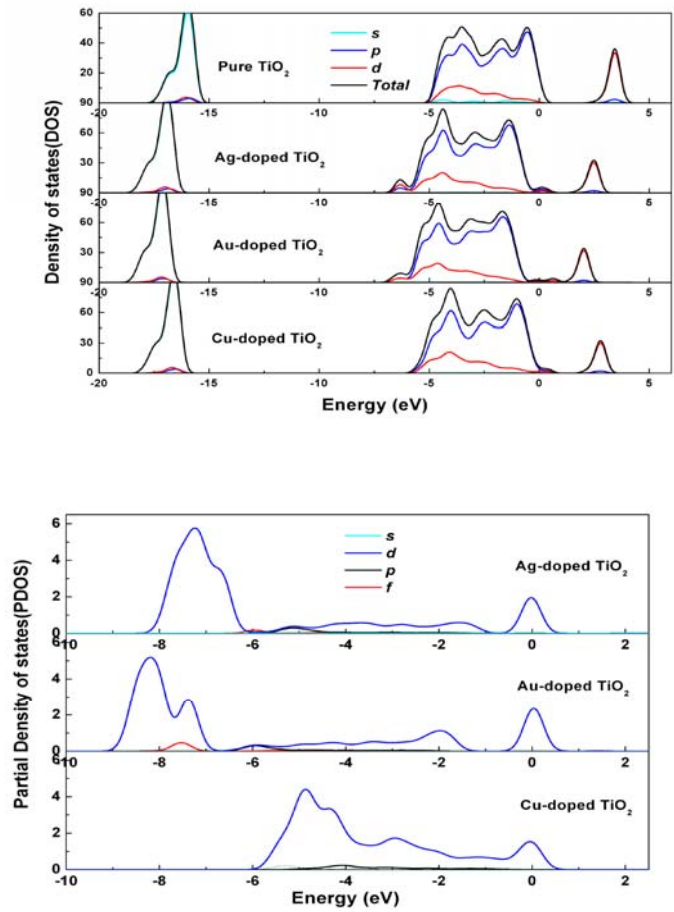

Fig.3 DOS of pure, Cu, Ag, and Au-doped TiO$_2$ (a), and PDOS of doping metals (b).

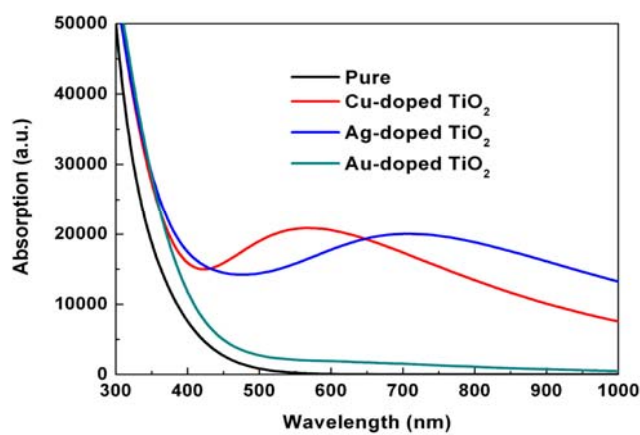

Fig.4 Optical absorptions of pure, Cu, Ag, and Au-doped $TiO_2$.

### 3.2 Optical properties of Cu, Ag, and Au-doped $TiO_2$

We carry out the optical absorption of Cu, Ag, and Au-doped $TiO_2$ under the scissor operation in Fig.4. Due to the underestimation of the band gap, it is difficult to obtain the exact optical band gap. This scissor operator can effectively describe the difference between the theoretical and experimental band gap values. When the experimental value is known, we can perform a band structure calculation to find the theoretical band gap. The relative position of the conduction to valence band is erroneous when the Kohn-Sham eigenvalues are used. In an attempt to fix this problem, inherent in DFT, we allow a rigid shift of the conduction levels to be consistent with the measured value of the band gap. In our calculation, the band gap of pure $TiO_2$ is 2.18 eV, which is underestimated by 1.02 eV compared with experimental data. Therefore, the 1.02 eV scissors operator has been used to evaluate the optical absorption. Compared with the pure

TiO$_2$, it is obvious that the sparsely doped Cu, Ag, Au atoms contribute to the in-gap impurity bands, while have few effects on the host's DOS. Therefore, the same scissor operator is applied to the doped TiO$_2$ systems. This method is effective for a variety of systems [37-40]. It can be observed that the doped TiO$_2$ shows the enhancement of visible light absorption. The Cu-doped TiO$_2$ presents the overall optical absorption in the range of 400-1000 nm, and the absorption center of Cu-doped TiO$_2$ locates at 570 nm. The visible light photoactivity of Cu-doped TiO$_2$ was also observed by other experiment reports [12, 14]. Colon *et al*. also reported that the Cu ion incorporated into TiO$_2$ could promote the visible photoactivity, and similar optical properties had also been observed [17]. Our result shows that the Cu ion doping can promote the visible optical absorption of TiO$_2$. Thus, if the preparation methods can be controlled appropriately, the Cu-doped TiO$_2$ can enhance the visible light response. In addition, the Ag-doped TiO$_2$ also shows evident visible absorption. The absorption center of Ag-doped TiO$_2$ locates at around 705 nm, and the absorption band edge shifts to the long wavelength region. The Au-doped TiO$_2$ presents the slight enhancement in the visible region compared with the pure TiO$_2$, Our optical absorption result agree well with that of the previous study [10]. It is necessary to point out that the similar dopant states (*d* electronic states) appear in the Ag and Au-doped TiO$_2$, however, they show the difference in the optical transitions.

The common accepted viewpoint is that dopant electronic states at the top of valence band can be controlled easily, and can promote visible photoactivity. Thus, the nonmetal N and C have been widely investigated [41, 42]. According to this viewpoint, we can primary conclude that the Cu-doped TiO$_2$ is a kind of very promising visible photocatalytic materials. Besides, the computational methods are also very important to the accuracy of doped-systems. The use of LDA (GGA)+U approach can decrease both hybridization and the energy of the valence-band top. For example, in the studied case of Fe-doped TiO$_2$, the LDA (GGA)+U can correct band gap to 2.8 eV [43]. Meanwhile, hybrid functional, such as PBE0 and HSE can correct band gap to 3.2 eV, but it is difficult to use in present systems. Therefore, the standard DFT is still feasible to investigate the relative changes of electronic structure after doping and is often rapid and effective.

**3.3 Optical transition mechanism of Cu, Ag, and Au-doped TiO$_2$**

To illustrate the optical transition mechanism in detail, the imaginary part of dielectric function has been given in Fig.5. It is well known that the imaginary part of dielectric function can describe the optical transition well. It can be found that the pure $TiO_2$ shows the optical transition of 4.93 eV, which is related to the intrinsic transition between O 2*p* states at the valence band and Ti 3*d* states at the conduction band. After doping, the intrinsic optical transition has a shift to the low energy range, which means that the band gaps are narrowed by doping Cu, Ag, and Au. The detailed optical transitions of Cu, Ag, and Au-doped $TiO_2$ are 4.67, 4.73, and 4.46 eV, respectively. Besides, the Cu-doped $TiO_2$ shows the optical transition of 1.78 eV, which is consistent with the result of optical absorption. This result indicates that the transition of 1.78 eV is due to the transition from Cu *d* electronic states to conduction band. The optical transition of Ag-doped $TiO_2$ locates at the 1.39 eV. Combining with the previous electronic states and absorption results, we can conclude that the 1.39 eV peak origins from the transition between the Ag 4*d* states in the band gap and conduction band. As for the Au-doped $TiO_2$, the visible optical transition of 2.19 eV is presented, which is attributed to the transition from the valence band to the midgap states of Au 5*d* states.

Finally, we present the optical transition mechanisms of Cu, Ag, and Au-doped $TiO_2$ in detail according to the electronic structures and optical properties. Combining with the band structure of Cu, Ag, and Au-doped $TiO_2$, optical transition mechanisms are illustrated in Fig.6. It can be observed that there are some midgap states for Cu, Ag and Au-doped $TiO_2$ in forbidden band. The Cu doping can induce the 3*d* states, which is near the uppermost valence band. It can be observed that two transitions can be occurred. The transition of E corresponds to the intrinsic transition. The transition of $E_1$ corresponds to the optical transition between Cu 3*d* electronic states and conduction band, which is consistent with the visible absorption around 570 nm. To the Ag-doped $TiO_2$, there are two transitions of E and $E_1$. The E also corresponds to the intrinsic transition, and the $E_1$ corresponds to the transition between Ag 4*d* states and conduction band, while transition from valence band to Ag midgap states can't take place. The transition of Au-doped $TiO_2$ is different from that of the Ag-doped $TiO_2$. and $E_1$ corresponds to the transition from valence band to Au 5*d* states. Li *et al*. also observed similar visible absorption, and assumed the transition mechanism of Au ion doped $TiO_2$ according to the experiment result, which showed that the visible absorption was attributed to the transition from Au energy level to conduction band. This mechanism of optical transition is not consistent with our conclusion, because the position of Au ion energy level has not directly obtained in their reports [10]. Our calculation confirms the position of Au ion energy level is near the conduction band. Therefore, we conceive that the visible absorption in our calculation is due to the optical transition between O 2$p$ and Au 5$d$ states.

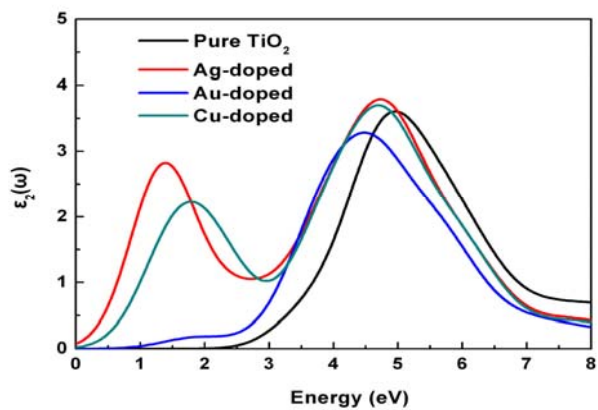

Fig.5 Imaginary parts of dielectric function of pure, Cu, Ag, and Au-doped TiO$_2$.

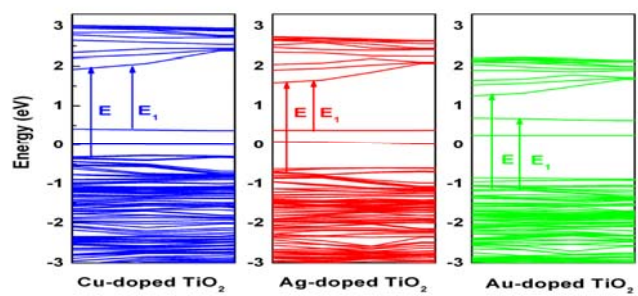

Fig.6 Optical transition mechanisms of Cu, Ag, and Au-doped TiO$_2$.

Now, we consider all these aspects including visible absorption and physical nature of three metal-doped TiO$_2$. The Cu, Ag, and Au doping cause the midgap states at the different positions in the forbidden gap, which is not the reason of band gap narrowing. Actually, these midgap states are localized states, not connecting with the valence band or conduction band, which can only account for the visible optical transition. The band gap narrowing of Cu, Ag, and Au-doped TiO$_2$ is attributed to the shift of uppermost valence band and lowest conduction band. The Cu-doped TiO$_2$ introduces visible optical transitions, namely Cu 3$d$-Ti 3$d$ states, while the optical transition between the O 2$p$ and Cu 3$d$ can not take place. In addition, the Ag and Au doping also create the middle states in the band gap, but the mechanism of optical transition is quite different. The Ag doping causes the visible optical transition between middle 4$d$ states and Ti 3$d$, and the optical transition between O 2*p* and middle 4*d* states is forbidden. As a contrast, the Au doping causes the visible optical transition between O 2*p* and middle Au 5*d* states, and the optical transition between Au 5*d* and Ti 3*d* states is forbidden. The present results clearly show that the middle stats introduced by doping have not always contributed to the visible optical absorption.

## 4. Conclusions

We perform the first-principles investigations of electronic structures and optical transitions of Cu, Ag, and Au-doped $TiO_2$. We find that the Cu doping can induce some doping states near the top of valance band, and thus enhance the visible absorption in the range of 400-1000 nm by Cu 3*d*-Ti 3*d* optical transition. Meanwhile, the Ag and Au doping induce the middle states in band gap of $TiO_2$. The Ag doping induces the visible light absorption in the range of 500-1000 nm due to the transition between Ag doping states and conduction band. Besides, the Au doping has also induced the visible light absorption enhancement due to the transition between valence band and Au doping states, but the absorption intensity is very low The present calculated results are helpful for understanding the physical nature of metal-doped $TiO_2$, and provide some useful information for the design of visible photocatalytic materials.

**Acknowledgments**

This work is supported by National Natural Science Foundation of China under Grant No. 11175128.